# Single Crystalline InGaAs Nanopillar Grown on Polysilicon with Dimensions beyond Substrate Grain Size Limit

Kar Wei Ng, Thai-Truong D. Tran, Wai Son Ko, Roger Chen, Fanglu Lu and Connie J. Chang-Hasnain[a]

*Department of Electrical Engineering and Computer Sciences, University of California at Berkeley, Berkeley, California 94720, USA*

**Abstract**

Monolithic integration of III-V optoelectronic devices with materials for various functionalities inexpensively is always desirable. Polysilicon (poly-Si) is an ideal platform because it is dopable and semi-conducting and can be deposited and patterned easily on a wide range of low cost substrates. However, the lack of crystalline coherency in poly-Si poses an immense challenge for high-quality epitaxial growth. In this work, we demonstrate, for the first time, direct growth of micron-sized InGaAs/GaAs nanopillars on polysilicon. Transmission electron microscopy shows that the micron-sized pillars are single-crystalline and single Wurzite-phase, far exceeding the substrate crystal grain size ~100nm. The high quality growth is enabled by the unique tapering geometry at the base of the nanostructure, which reduces the effective InGaAs/Si contact area to < 40 nm in diameter. The small footprint not only reduces stress due to lattice mismatch but also prevents the nanopillar from nucleating on multiple Si crystal grains. This relaxes the grain size requirement for poly-Si, potentially reducing the cost for poly-Si deposition. Lasing is achieved in the as-grown pillars under optical pumping, attesting their excellent crystalline and optical quality. These promising results open up a pathway for low-cost synergy of optoelectronics with other technologies such as CMOS integrated circuits, sensing, nanofluidics, thin film transistor display, photovoltaics, etc.

---

[a] Electronic mail: cch@berkeley.edu.

# Manuscript Text

III-V compound materials have been an important building block in optoelectronics technology due to their direct bandgaps leading to excellent optical and electrical properties. Conventionally, high-quality epitaxial growth requires the use of III-V substrates, which are usually expensive and fragile. Heterogeneous growth of III-V materials on substrates with lower cost is thus highly desirable. Among various low-cost substrates, polycrystalline silicon (poly-Si) is an excellent choice because of its unique semi-conducting property to facilitate doping into both p- and n-type for various semiconducting device functionalities. Moreover, poly-Si can be deposited and patterned easily on a great variety of carrier substrates, such as glass, polymer, metal, and oxidized silicon substrates, just to name a few. With these unparalleled properties, poly-Si has become one of the most important electronic materials used in a vast variety of applications including CMOS transistor, sensing, nanofluidics, thin film transistor (TFT) in displays, photovoltaics, etc.[1-5]

Being able to grow high-quality III-V structures on poly-Si could open up a new pathway for low-cost monolithic integrations of optoelectronics with these many applications. In particular, single-crystalline structures that are micron-sized are critical to reduce surface-to-volume ratios. However, the growth of single crystalline III-V structures on poly-Si remains to be challenging due to the extremely short-range crystalline coherency in poly-Si. There have been previous reports of III-V semiconductor nanowire growth on low-cost amorphous and poly-crystalline substrates [6-9]. In the former case, only nanowires with diameter as small as 50 nm are shown to have high crystal purity. Moreover, the amorphous substrate is electrically insulating, making electrical contact to the nanowire difficult. On poly-crystalline substrates, the diameter of the nanowire is limited to well below 200 nm by the grain size of the substrate to ensure that the nanostructure is not nucleating from multiple substrate crystal grains. Increasing the size of the nanostructure while maintaining high crystal purity thus is a major challenge.

In this paper, we report the growth of single crystalline $In_{0.16}Ga_{0.8}As$/GaAs core-shell nanopillar on poly-Si via MOCVD at a low temperature of 400°C. The base diameter of the nanopillars is over 800 nm, which far exceeds the grain size of the substrate (~100 nm). Nanopillars are randomly oriented due to the polycrystalline nature of the substrate. A density as high as $10^8$ cm$^{-2}$ is achieved. With the use of focused ion beam (FIB) and high-resolution transmission electron microscopy (HRTEM), we cut through the center of nanopillars and examine the InGaAs/poly-Si interface. The InGaAs nanopillar is observed to stem directly on a polycrystalline substrate. In addition, the bulk of the pillars is single crystalline with pure wurtzite (WZ) crystal phase. The key to this high-quality mismatched growth is the inverse-cone shape at the base of the nanostructure. This feature reduces the footprint of the nanostructure down to below 40 nm in diameter and ensures that III-V material seeds on a single Si crystal grain. The vertical growth direction [0001] is found to align with Si [110] rather than the conventional Si [111] direction.

This crystalline alignment leads to discrepancy in lattice arrangement across the hetero-interface, facilitating the formation of misfit dislocations which relax stress due to 5.3% lattice mismatch. Photoluminescence is utilized to further verify the crystalline and optical quality of the nanostructures. Lasing is achieved upon optical pump with the *as-grown* nanopillars, attesting their excellent crystalline and optical quality. This is the first report of as-grown lasers on poly-silicon. These remarkable results are first proof that these nanostructures can potentially be used as high performance devices on top of poly-Si. The study presented here underscores a new methodology for inexpensive integration of materials for various functionalities with high-quality optoelectronic devices.

Figure 1 depicts the schematic of a typical sample and scanning electron micrograph (SEM) images of InGaAs/GaAs nanopillars directly grown on top of an 800-nm-thick polysilicon layer. The nanopillar growth was carried out by low-temperature (400 $^o$C) MOCVD. The poly-Si layer was deposited on an oxidized silicon substrate by low-pressure chemical vapor deposition. Prior to the nanopillar growth, the substrate was cleaned and deoxidized with acetone, methanol and buffered oxide etchant, followed by surface mechanical roughening. Other growth parameters are similar to those used for growth on silicon substrate in our previous work.[10]

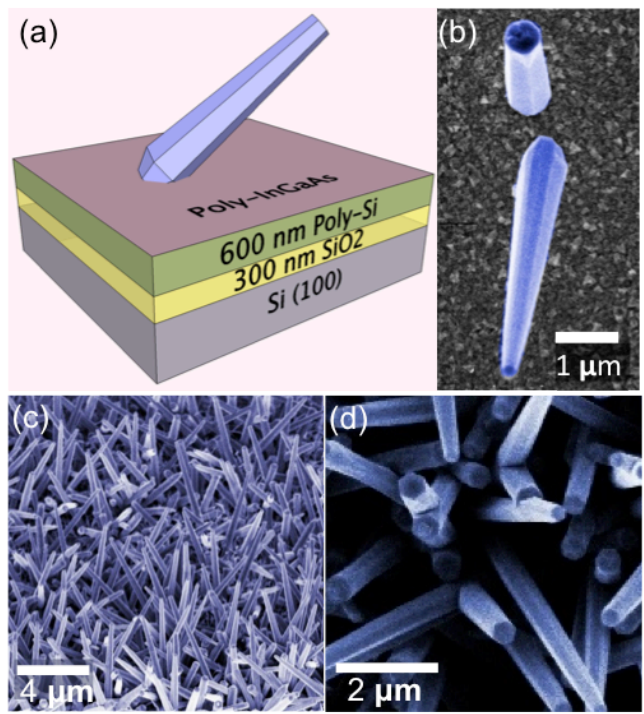

Figure 1. (a) Schematic diagram illustrating the growth of InGaAs/GaAs nanopillar on polysilicon. (b) Top view SEM images showing two typical standalone nanopillars. (c) Ensemble of nanopillars with random orientations due to the polycrystalline nature of the substrate. Density can be as high as $10^8$ cm$^{-2}$. (d) Close-up image of the forest. Hexagonal shape is well preserved. As the density is very high, some of the pillars cross each other during the growth.

We observed some variation in nanopillar density over 1"-diameter substrate, most likely due to V/III ratio, temperature and surface non-uniformity. In the low-density regions, individual nanopillars can be identified distinctively, as shown in the scanning electron microscope (SEM) image in Figure 1(b). The nanopillars are found to orient randomly as a result of the polycrystalline nature of the substrate. Despite being grown on a polycrystalline substrate, the nanopillars possess clear facets, which are first indications that they are crystalline structures. In an 80-min growth, the base diameter and height can reach ~ 800 nm and 4 μm, respectively. On the other hand, in the high-density regions, nanopillars grow into an ensemble with a density as high as $10^8$ cm$^{-2}$, as illustrated in Fig 1(c). These densely packed nanostructures have much smaller diameters, ~400 nm, and are randomly oriented such that some of the nanopillars are actually crossing one another, which is a signature of the characteristic core-shell growth mode. In spite of the crossed growth, the nanopillars do exhibit distinctive hexagonal cross-sections and smooth sidewalls, as seen in the close-up image in Figure 1(d). The high density and randomly aligned nanopillar ensemble can trap and scatter light effectively, which make them ideal candidates for light sensing and photovoltaic applications.[11]

Given the lack of long-range crystalline coherence in poly-silicon, it is crucial to find out if the nano- or micro- pillars are actually of good quality. Here, we study this topic extensively with high resolution TEM (HRTEM) and scanning transmission electron microscopy (STEM). We examine the very center of the nanopillar by cutting across the pillar with FIB and expose the ($2\bar{1}\bar{1}0$) facets, or a-planes, to clearly disclose the distinction between wurtzite and zincblende (ZB) phases. Details of sample preparation can be found in our previous work.[12] More than ten nanopillars growing along different directions were examined in this work to obtain a general qualification of crystalline quality.

Figures 2 (a) and (b) show a high angle annular dark field (HAADF) STEM image and schematic illustration of a close-to-upright nanopillar grown on poly-Si. Poly-Si possesses a rather rough surface being deposited on an oxidized silicon substrate. Despite of the roughness, InGaAs is observed to stem directly from poly-Si without any void in-between, as seen in the bright field TEM image of the pillar base in Figure 2(c). As the substrate is polycrystalline, one would expect any crystal nucleating directly on top should have mediocre quality. Surprisingly, the bulk of the 650-nm thick pillar is single crystalline, as attested by the clear diffraction pattern in the inset of Figure 2(d). The unique zigzag lattice arrangement in the HRTEM image in Fig 2(d) further confirms that the crystal is in pure WZ phase.

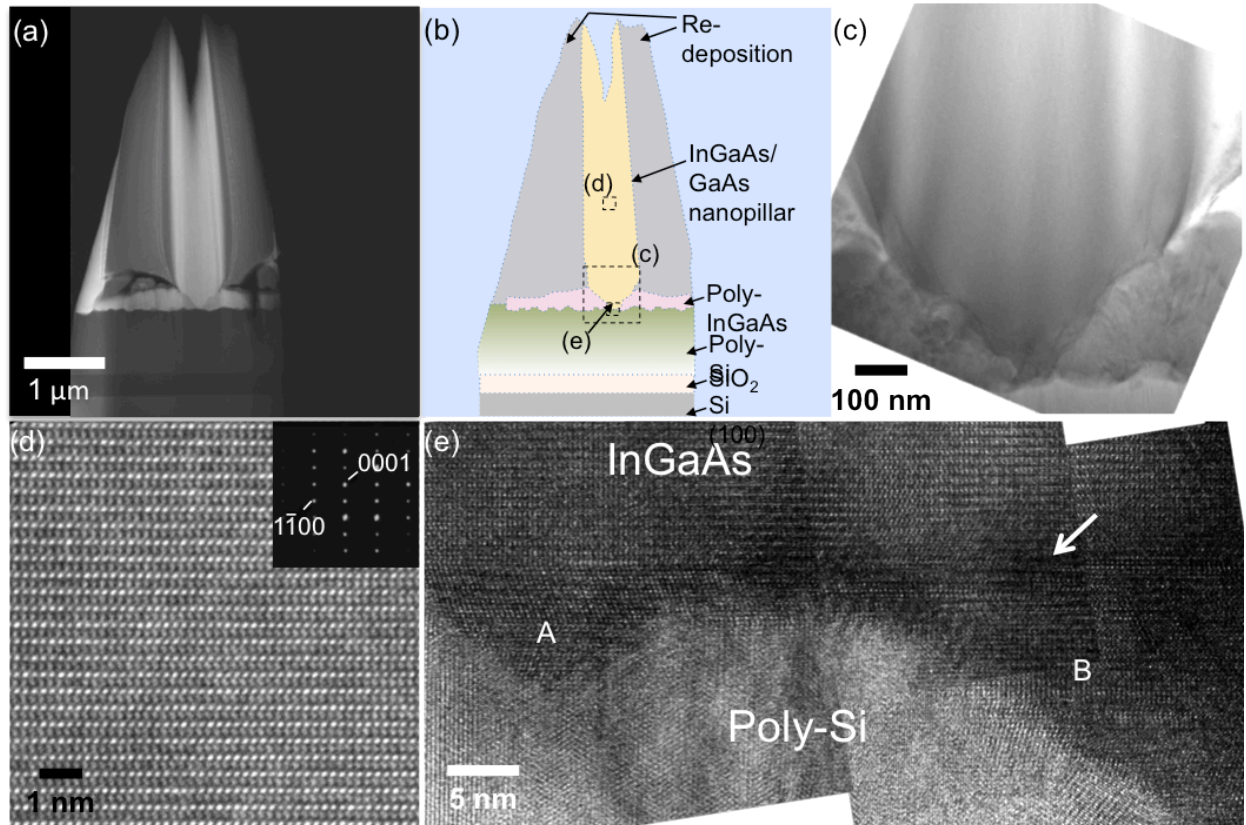

Figure 2 (a) HAADF-STEM image of a needle grown on poly-Si. (b) Schematic illustration of the structure in (a). (c) TEM image at the base of the nanoneedle. The base tapers into a small footprint on the poly-Si as a result of confined growth. Stacking faults are absent throughout the entire needle. (d) HRTEM image of InGaAs along [$2\bar{1}\bar{1}0$]. The clear zig-zag lattice arrangement and diffraction pattern in the inset show the pure wurtzite crystal phase of the structure. (e) HRTEM image at the exact InGaAs/Poly-Si interface. In spite of the waviness of the interface, the crystal turns into perfectly coherent wurtzite phase within 10 nm from the interface.

Surrounding the nanopillar, InGaAs evolves into polycrystalline thin film. The same poly-InGaAs layer is also observed in the growth of (In)GaAs nanostructures on (111)-Si and sapphire substrates,[13-14] indicating that the formation of such layer is likely due to the low growth temperature and high lattice mismatch rather than crystallinity of the substrate. This polycrystalline layer grows simultaneously with the single crystalline structure throughout the entire growth process and restricts the pillar base from expanding, thus resulting in inverse tapering at the root, as seen in Figure 2(a) and (c). Consequently, the diameter of the bulk can expand to sub-micron (~ 650 nm in this particular TEM sample) while the footprint remains to be < 40 nm. The diameter of nucleation site, and thus the nanowire diameter, has to be smaller than the Si grain size or otherwise III-V would nucleate on multiple grains and eventually evolve into polycrystalline structures.[7] The cone-shaped base presented in this work thus allows the growth of single crystalline sub-micron sized structures on polycrystalline platforms with grain sizes as small as 40 nm.

The inverse cone geometry at the pillar base reduces stress developed in the nanostructure due to over 5% lattice mismatch. In InGaAs nanopillar grown on (111)-Si, the remaining stress is relaxed via formation of stacking disorders, which are well confined within the inversely tapered region.[12] While this is the case as well for some of the nanopillars grown on poly-Si, there exists nanopillars on poly-Si in which stacking fault is completely absent in the inverse-cone, like the one shown in Figure 2. To understand how misfit stress can be relaxed, it is essential to study the InGaAs/poly-Si interface. Figure 2(e) shows an HRTEM image along [2$\bar{1}\bar{1}$0] zone axis. The poly-Si surface is rough even in nanoscopic scale, resulting in random morphologies along the hetero-interface. We believe that the roughness creates a lot of dangling bonds over the surface and lowers the energy barrier for III-V nucleation on Si, thus enabling the high-density growth observed. The roughness along [2$\bar{1}\bar{1}$0], i.e. the direction pointing into the paper, causes overlapping of different lattices and this leads to a blurry hetero-interface in Figure 2(e). Despite the nanoscopic roughness, InGaAs is observed to grow directly on top of the poly-Si layer without any amorphous material in between. The bottommost 3~10 nm is a transition regime composed of ZB/WZ mixed layers. Within the transition regime, we note that the crystal in zone A is in ZB phase while that in zone B is in WZ phase. This is rather subtle as the crystal phase is usually coherent along the lateral direction, i.e. perpendicular to [0001]. One possible reason is that InGaAs nucleated at the two sites separately. Depending on the local surface morphology of poly-Si, InGaAs evolved into either ZB or WZ phase to minimize the interface energy, hence resulting in the observed lateral crystal incoherency. Lattice disordering is observed at where the grains meet, as indicated by an arrow in Figure 2(e). These defects do not propagate upward because the crystal at the base expands in the lateral direction only as a result of the characteristic core-shell growth mode. InGaAs then evolves into pure WZ phase above the transition and this continues up to the tip of the nanopillar. With the unique core-shell growth, high quality single-crystalline nanopillar can be obtained even on a bumpy poly-Si surface.

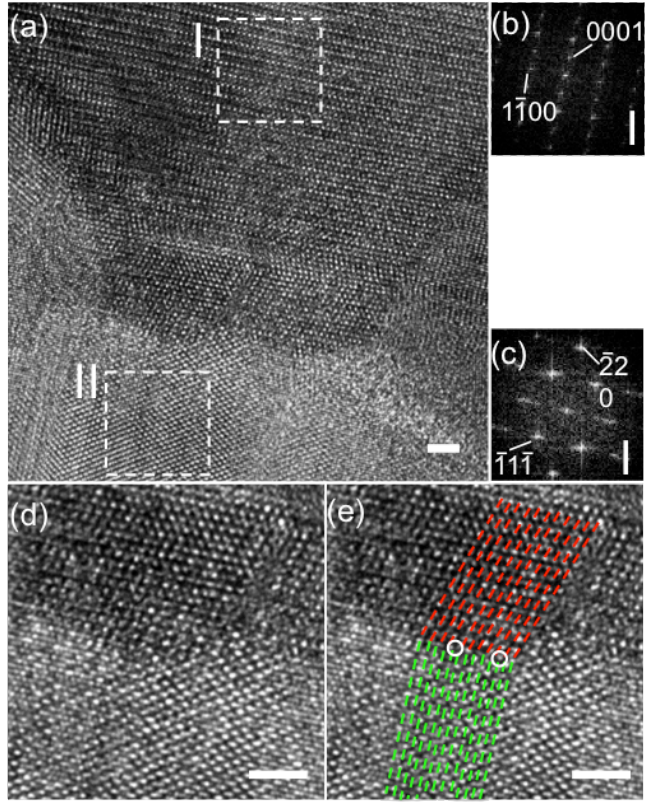

Figure 3 (a) HRTEM of an InGaAs/poly-Si interface along $[2\bar{1}\bar{1}0]$ zone axis with the exact contact as small as 15 nm in diameter. (b) & (c) FFTs of region I (WZ phase InGaAs) and region II (ZB phase poly-Si), respectively. InGaAs [0001] is observed to be aligned to Si [110]. (d) Magnified image of the InGaAs/poly-Si interface shown in (a). (e) The same image in (d) with read and green dotted lines as eyeguides for InGaAs $(1\bar{1}1)$ and Si (001) planes. Interfacial misfit dislocations are indicated by white circles. All scale bars indicate 2 nm.

Although the nanopillar shown in Figure 2 provides much information on the influence of nanoscopic roughness on nucleation, the hetero-interface is undesirably blurry and thus fine details cannot be interpreted clearly. To study the crystalline relation between III-V and the substrate, HRTEM of the hetero-interface of another nanopillar is presented in Figure 3(a). The footprint of this particular nanopillar is only 15 nm in diameter. Below the pure WZ phase crystal is a 5nm transition region composed of pure ZB phase crystal. Fast Fourier transforms (FFTs) of the lattices in region I (i.e. WZ phase InGaAs) and region II (i.e. ZB phase poly-Si) are shown in Figure 3(b) and (c), respectively. We note that FFT of region II is a bit hazy due to overlap of crystal lattices with slightly different orientations in the poly-Si substrate. Nevertheless, it can be observed that InGaAs [0001] (or [111] in the bottommost ZB transition region) is aligned to Si [110], i.e. InGaAs (111) in the transition region is in direct contact with Si (110) at the hetero-interface. In fact, the same crystalline relation is also observed in zone A of Figure 2(e). Compared to Si (111), Si (110) has considerably higher surface energy and is thus more favorable for III-V nucleation.[15] However, this crystalline dependence is rather uncommon as (111) and (110) planes are quite different in lattice arrangements. Such discrepancy in lattice arrangement is actually beneficial as it favors the formation of dislocations and dangling bonds at the InGaAs/Si interface, as illustrated in Figures 3(d) and (e). These interfacial defects can effectively relax most of the misfit stress between the substrate and the epitaxial material and this is also observed in the hetero-integration of other material systems.[16] The unconventional crystalline relation, together with the inverse cone geometry at the base, minimize compressive stress induced by over 5% lattice mismatch and enables of the growth of stacking disorder free single crystalline III-V nanopillar on poly-Si.

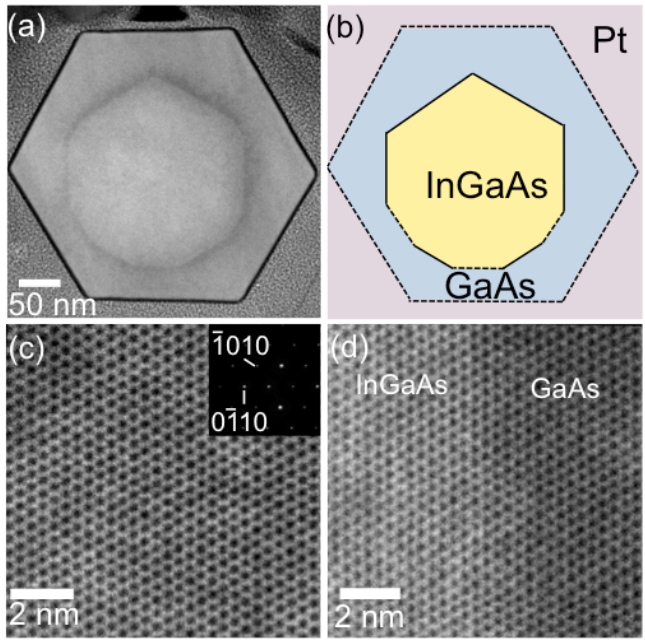

Figure 4 (a) HAADF-STEM image of an InGaAs/GaAs nanoneedle along c-axis. (b) Schematic diagram of the structure in (a). The dotted and solid lines represent a-planes and m-planes, respectively. (c) HR-STEM image of InGaAs. The hollow hexagonal lattice arrangement reveals the WZ nature of the crystal. Inset is the selective area diffraction pattern along c-axis. (d) HR-STEM image of an InGaAs/GaAs interface. GaAs shell grows seamlessly on $In_{0.2}Ga_{0.8}As$ shell.

Similar to nanoneedles grown on sapphire and crystalline Si substrates, InGaAs/GaAs nanopillars grow in a core-shell manner on top of poly-Si. The characteristic core-shell geometry facilitates elastic misfit stress relaxation, enabling the growth of single crystalline GaAs with thickness far beyond the thin film critical limit.[17] To visualize the core-shell growth mode, we cut across a nanopillar horizontally to expose the c-plane, i.e. (0001) plane. HAADF-STEM image and the corresponding schematic illustration of the cross section are shown in Figure 4(a) and (b), respectively. A clear boundary can be seen at the InGaAs/GaAs boundary. While GaAs shell is perfectly hexagonal with sidewalls composed of $\{11\bar{2}0\}$ (i.e. a-planes), InGaAs core is a nonagon bounded by six $\{1\bar{1}00\}$ (i.e. m-planes) and three a-planes. We attribute this asymmetry to non-uniform III-adatom supply around the nanostructure. As the density can be as high as $10^8$ cm$^{-2}$, significant competitions of adatoms occur among neighboring nanopillars. However, since the nanostructures have random orientations, one side of the pillar may experience more crowding effect than the other side, resulting in inhomogeneous growth rate. In the structure shown in Figure 4, GaAs is thinner in the lower half of the cross-section, which is likely to be a consequence of asymmetric resource competition. Furthermore, there are studies showing that the formation of m-planes is more energy favorable under high group-III supply while a-planes appear when group III supply is low.[18] A mix of a- and m-planes in the lower half of the core further confirm that group III supply is low on that side the nanopillar. As a result, the uneven group III adatom supply give rise to asymmetry in pillar shape and this can impact the optical cavity properties of the nanostructures.

Although there is an irregularity in the shape of the core, the quality of the crystal is not compromised. Figure 4(c) shows a HR-STEM image of InGaAs core along [0001]. The lattice assumes a honeycomb pattern, which is a distinctive feature of WZ phase crystal. The clear diffraction pattern in the inset further proves the crystal purity. At the InGaAs/GaAs interface, the honeycomb pattern continues seamlessly across the border without any noticeable misfit defects, as illustrated in Figure 4(d). This smooth transition across the hetero-interface is observed in the entire structure, independent of whether the border is an a-plane or m-plane. This observation confirms that the core-shell geometry indeed enables high-quality mismatched growth of layers exceeding the thin film critical thickness, no matter what substrate is being used.

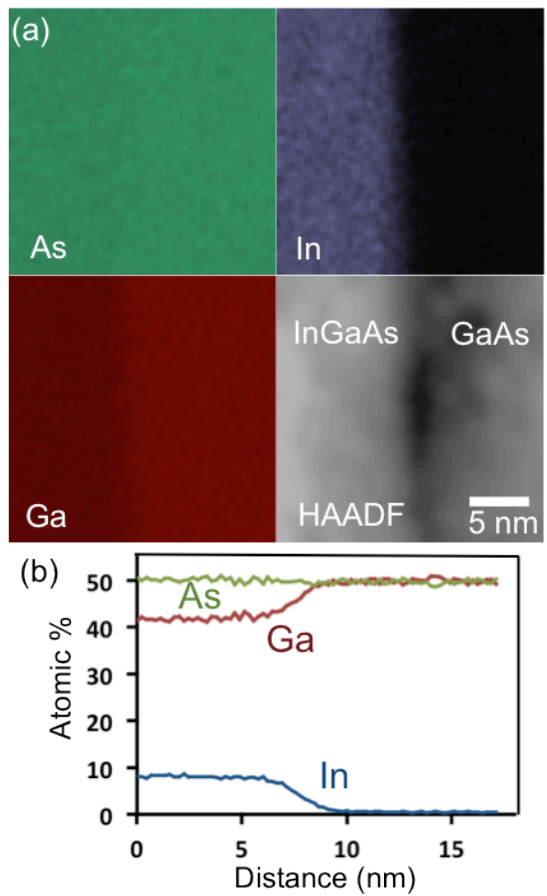

Figure 5 (a) Elemental EDS maps and HAADF-STEM image of an InGaAs/GaAs interface. (b) Element concentration line scan profile across the hetero-interface shown in (a). Indium cuts off sharply at the border with a transition as small as 2 nm.

Interface sharpness has significant impact on the optical properties of a nanostructure. We performed energy dispersive x-ray spectroscopy (EDS) to understand the compositional variation at the InGaAs/GaAs interface. Figure 5(a) shows the elemental maps of Ga, In and As of the interface depicted in the HAADF-STEM image. While As shows a uniform distribution over the entire area, a sharp cut off is observed in the In map at the hetero-interface. The Ga map also shows a subtle intensity change through the boundary, indicating the rise of Ga abundance from $In_{0.16}Ga_{0.84}As$ to GaAs. Figure 5(b) presents the quantitative profiles of the three elements in the area depicted in Figure 5(a). Indium composition drops abruptly at the hetero-interface with a transition of around 2 nm. We anticipate that the actual transition region is even thinner as the EDS resolution is limited by electron scattering in the sample. Nevertheless, this is one of the sharpest interfaces as characterized by EDS, to the best of our knowledge. These observations attest to the abruptness of the InGaAs/GaAs interface, which gives rise to excellent crystal quality of the nanostructures.

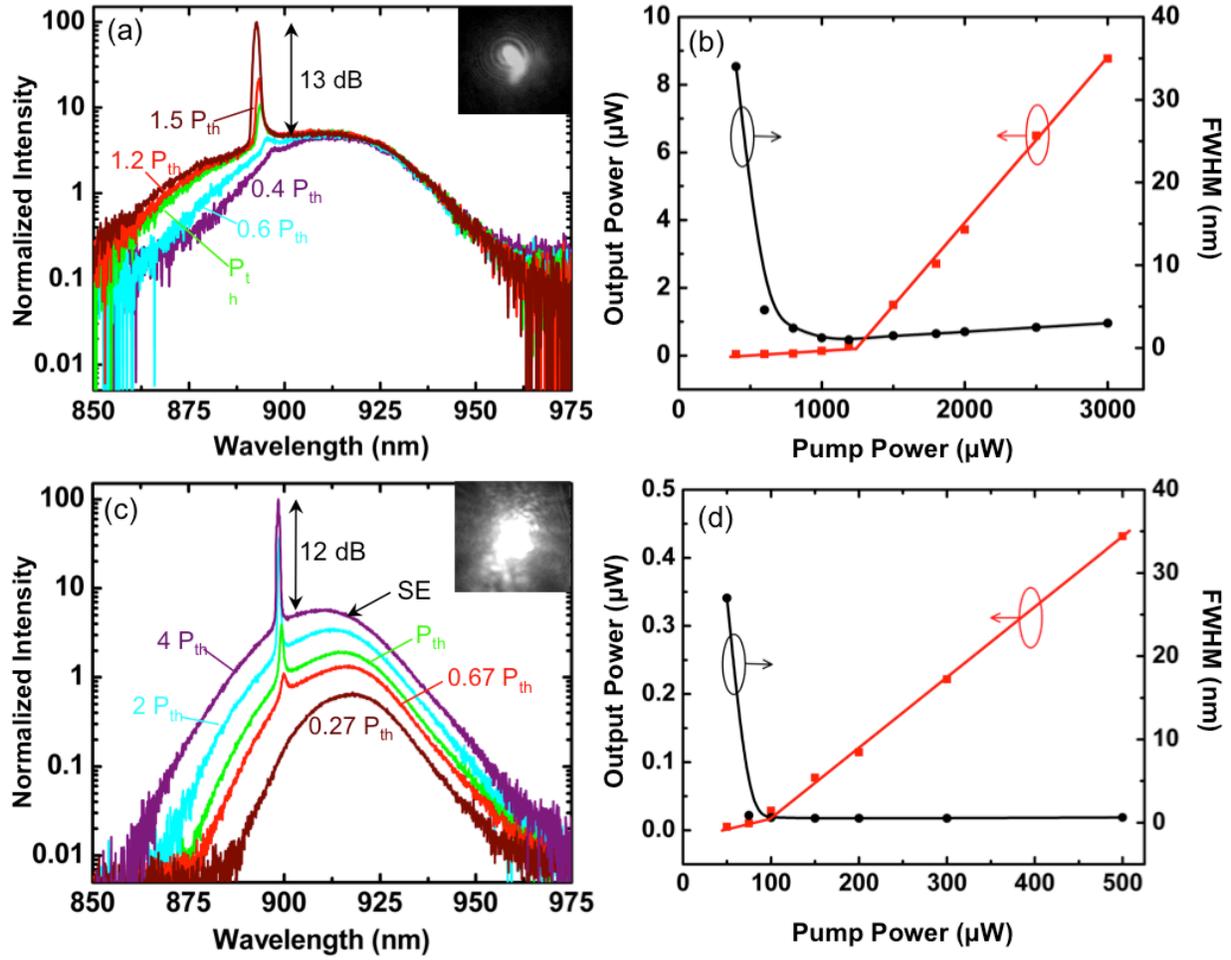

Figure 6. **Nanopillar Lasers:** Emission spectra nanopillar under different pump powers obtained from (a) standalone nanopillar (c) nanopillar ensemble. Sideband suppression ratios (SSR) of ~ 12 dB is observed in both cases. The pump power required for the same SSR is much higher in nanopillar ensemble than standalone ones due to spontaneous emissions from the neighboring pillars. Insets show speckle patterns in the near-field upon lasing. Light emission power and line width plotted as a function of pump power from (b) standalone nanopillar (d) nanopillar ensemble. Clear threshold behavior is observed. Line width reduces by a factor of ~ 30 near and above the laser threshold in both cases.

With the excellent crystalline quality, lasing is achieved with both individual and ensemble InGaAs/GaAs nanopillars under optical pump at 4K with Ti:sapphire laser. Figure 6(a) shows the emission spectra of a standalone nanopillar under various pump powers. At low pump levels, spontaneous emission is observed with a peak wavelength near 920 nm. As the excitation power increases, a cavity mode emerges at around 895 nm, which finally reaches laser oscillation. Near field optical image for a standalone nanopillar laser above threshold is shown in inset. Lasing is attributed to a helically propagating mode in which light spirals along the well-faceted hexagonal nanopillar.[10] Although there is a detuning between gain and cavity mode, a 13dB side-mode suppression ratio is observed at 1.5 times threshold pump power ($P_{th}$). Figure 6(b) shows the output power and spectra linewidth as a function of pump power. Clear threshold behavior is observed in light output power and linewidth narrowing.

Lasing is achieved with nanopillar ensemble for the first time in arsenide material system, as seen in Figure 6(c). Unlike single pillar laser, spontaneous emission increases linearly with pump power even above lasing threshold. As the pillar density is high in the ensemble, multiple pillars are excited at the same time upon optical pump. The nanopillars in the ensemble have much smaller diameters and, hence, lower optical quality factors than those of the standalone pillars. Lasing is attributed to scattered feedback from multiple pillars, similar to the random lasing effect.[19-20] As a result, spontaneous emission is not clamped above threshold. Side mode suppression ratio reaches 12 dB at a rather high pump power level of $4P_{th}$, considerably higher than standalone nanopillar laser. However, the lasing threshold is also significantly lower. Near field optical image of the ensemble laser is shown in the inset. Significant coherent speckle is observed due to scattering from multiple pillars, which is also distinct from that of standalone pillar laser. Clear threshold behavior is observed in the pump power dependence of light output power, as seen in Figure 6(d). Linewidth reduces by over 30 times near and above the threshold, attesting the wavelength coherency of the laser emission. These exciting results reveal the excellent optical properties of these nanostructures and underscore their potential usefulness as efficient optical components that can be integrated inexpensively with various substrates or carriers.

In summary, we demonstrated the growth of single crystalline InGaAs/GaAs pillars on polycrystalline silicon. The pillar base diameter approaches a micron, which is far greater than the substrate crystal grain size. With the inverse tapering at the base, the effective footprint of the nanostructure is reduced to below 40 nm in diameter. The small contact area prevents the III-V crystal from nucleating on top of multiple grains in the substrate and minimizes stress in the system due to lattice mismatch. In fact, the small footprint lowers the grain size requirement on the poly-Si film or substrate and this can potentially lower the production cost. The special crystal alignment between InGaAs and Si facilitates formation of misfit dislocations at the interface, thus allowing the growth of stacking fault free III-V crystal on poly-Si. Lasing is demonstrated upon optical pump, attesting the excellent crystal and optical quality of the as-grown nanostructures. These properties reveal the potential of these nanostructures to be used for various optoelectronic applications. The results presented in this work open up a pathway for low-cost synergy of III-V based optoelectronics with many technologies, e.g. TFT and nanofluidics, via poly-Si as a bridging platform.

**Acknowledgements**

This work was supported by U.S. DOE Sunshot Program (DE-EE0005316), DoD NSSEFF Fellowship (N00244-09-1-0013 and N00244-09-1-0080), and the Center for Energy Efficient

Electronics Science (NSF Award 0939514). We acknowledge support of the National Center for Electron Microscopy, LBL, which is supported by the U.S. DOE (DE-AC02-05CH1123).

**References**

1. Xiong, Z.; Liu, H.; Zhu, C.; Sin, J.K.O. A new polysilicon CMOS self-aligned double-gate TFT technology. *IEEE T Electron Dev*. **2005**, *52*, 2629-2633.

2. Obermeier, E.; Kopystynski, P. Polysilicon As a Material for Microsensor Applications. *Sensor Actuat A-Phys*. **1992**, *30*, 149-155.

3. Kutchoukova, V.G.; Pakulaa, L.; Parikesitb, G.O.F.; Garinib, Y.; Nanvera, L.K.; Bosschea, A. Fabrication of nanofluidic devices in glass with polysilicon electrodes. *Sensor Actuat A-Phys*. **2005**, *123-124*, 602-607.

4. Stewart, M.; Howell, R.S.; Pires, L.; Hatalis, M.K. Polysilicon TFT technology for active matrix OLED displays. *IEEE T Electron Dev*. **2001**, *48*, 845-851.

5. Fossum, J.G.; Lindholm, F.A. Theory of grain-boundary and intragrain recombination currents in polysilicon p-n-junction solar cells. *IEEE T Electron Dev*. **1980**, *27*, 692-700.

6. Dhaka, V.; Haggren, T.; Jussila, H.; Jiang, H.; Kauppinen, E.; Huhtio, T.; Sopanen, M.; Lipsanen, H. High Quality GaAs Nanowires Grown on Glass Substrates. *Nano Lett.* **2012**, *12*, 1912–1918.

7. Ikejiri, K.; Ishizaka, F.; Tomioka, K.; Fukui, T. GaAs nanowire growth on polycrystalline silicon thin films using selective-area MOVPE. *Nanotechnology* **2013**, *24*, 115304.

8. Chandrasekarana, H.; Sunkara, M. K. Growth of Gallium Nitride Textured Films and Nanowires on Polycrystalline Substrates at sub-Atmospheric Pressures. *Mat. Res. Soc. Symp. Proc.* **2002**, *693*, I3.30.

9. Jacobs, R. N.; Salamanca-Riba, L.; He, M.; Harris, G. L.; Zhou, P.; Mohammad, S. N.; Halpern, J. B. Structural Characterization of GaN Nanowires Fabricated via Direct Reaction of Ga Vapor and Ammonia. *Mat. Res. Soc. Symp. Proc.* **2001,** *675*, W9.4.

10. Chen, R.; Tran, T. T. D.; Ng, K. W.; Ko, W. S.; Chuang, L. C.; Sedgwick, F. G. Chang-Hasnain, C. J. Nanolasers Grown on Silicon. *Nat. Photonics* **2011**, *5*, 170-175.

11. Miller, O. D.; Yablonovitch, E.; Kurtz, S. R. Strong Internal and External Luminescence as Solar Cells Approach the Shockley–Queisser Limit. *IEEE Journal of Photovoltaics* **2012,** *2*, 303-311.

12. Ng, K. W.; Ko, W. S.; Tran, T. T. D.; Chen, R.; Nazarenko, M. V.; Lu, F.; Dubrovskii, V. G.; Kamp, M.; Forchel, A.; Chang-Hasnain, C. J. Unconventional Growth Mechanism for Monolithic Integration of III–V on Silicon. *ACS Nano* **2013**, *7*, 100–107.

13. Chuang, L. C.; Moewe, M.; Ng, K. W.; Tran, T. T. D.; Crankshaw, S.; Chen, R.; Ko, W. S.; Chang-Hasnain, C. J. GaAs Nanoneedles Grown on Sapphire. *Appl. Phys. Lett.* **2011**, *98*, 123101.

14. Moewe, M.; Chuang, L. C.; Crankshaw, S.; Chase, C.; Chang-Hasnain, C. J. Atomically sharp catalyst-free wurtzite GaAs/AlGaAs nanoneedles grown on silicon. *Appl. Phys. Lett.* **2008,** *93*, 23116.

15. Messmer, C.; Bilello, J. C. The surface energy of Si, GaAs, and GaP. *J. Appl. Phys.* **1981,** *52*, 4623.

16. Jallipalli, A.; Balakrishnan, G.; Huang, S. H.; Rotter, T. J.; Nunna, K.; Liang, B. L.; Dawson, L. R.; Huffaker, D. L. Structural Analysis of Highly Relaxed GaSb Grown on GaAs Substrates with Periodic Interfacial Array of 90° Misfit Dislocations. *Nanoscale Research Letters* **2009**, *4*, 1458-1462.

17. Nazarenko, M. V.; Sibirev, N. V.; Ng, K. W.; Ren, F.; Ko, W. S.; Dubrovskii, V. G.; Chang-Hasnain, C. J. Elastic energy relaxation and critical thickness for plastic deformation in the core-shell InGaAs/GaAs nanopillars. *J. Appl. Phys.* **2013,** *113*, 104311.

18. Yamashita, T.; Akiyama, T.; Nakamura, K.; Ito, T. Growth of side facets in InP nanowires: First-principles-based approach. *Surf. Sci.* **2013,** *609*, 207-214.

19. Liu, X. Y.; Shan, C. X.; Wang, S. P.; Zhang, Z. Z.; Shen, D. Z. Electrically pumped random lasers fabricated from ZnO nanowire arrays. *Nanoscale*, **2012**, *4*, 2843.

20. Chen, R.; Utama, M. I. B.; Peng, Z.; Peng, B.; Xiong, Q.; Sun H. Excitonic Properties and Near-Infrared Coherent Random Lasing in Vertically Aligned CdSe Nanowires. *Adv. Mater.* **2011,** *23*, 1404–1408.